%% file: lyman_alpha.tex
\documentclass[useAMS,usenatbib]{mn2e}
\usepackage{Times}
\usepackage{psfrag}
\usepackage{pspicture}
\usepackage{graphicx}
\usepackage{lscape}
\newcommand\lsim{~\lower.5ex\hbox{$\buildrel < \over \sim$}~}
\newcommand\gsim{~\lower.5ex\hbox{$\buildrel > \over \sim$}~}

\title[A 10 deg$^2$ Lyman-$\bf \alpha$ survey at $z=8.8$]{A 10 deg$^2$ Lyman-$\bf \alpha$ survey at $\bf z=8.8$ with spectroscopic follow-up: strong constraints on the LF and implications for other surveys\thanks{Based on observations obtained with WIRCam on the CFHT, OPTICON programme 2011BA016, 2012A019 and 2012BA022.} }
\author[J.J.A. Matthee et al.]{Jorryt J.A. Matthee$^{1}$\thanks{E-mail: matthee@strw.leidenuniv.nl}, David Sobral$^{1}$, A.M. Swinbank$^{2}$, Ian Smail$^{2}$, P. N. Best$^{3}$, \newauthor Jae-Woo Kim$^{4}$, Marijn Franx$^{1}$, Bo Milvang-Jensen$^{5}$, Johan Fynbo$^{5}$\\
$^{1}$ Leiden Observatory, Leiden University, P.O.\ Box 9513, NL-2300 RA Leiden, The Netherlands\\
$^{2}$ Institute for Computational Cosmology, Durham University, South Road, Durham, DH1 3LE, UK\\
$^{3}$ SUPA, Institute for Astronomy, Royal Observatory of Edinburgh, Blackford Hill, Edinburgh, EH9 3HJ, UK \\
$^{4}$ Center for the Exploration of the Origin of the Universe, Department of Physics and Astronomy, Seoul National University, Seoul, Korea\\
$^{5}$ Dark Cosmology Centre, Niels Bohr Institute, University of Copenhagen, Juliane Maries Vej 30, 2100 Copenhagen, Denmark}
\begin{document}


\pagerange{\pageref{firstpage}--\pageref{lastpage}} \pubyear{2013}

\maketitle

\label{firstpage}

\begin{abstract}
Candidate galaxies at redshifts of $z \sim 10$ are now being found in extremely deep surveys, probing very small areas. As a consequence, candidates are very faint, making spectroscopic confirmation practically impossible. In order to overcome such limitations, we have undertaken the CF-HiZELS survey, which is a large area, medium depth near infrared narrow-band survey targeted at $z=8.8$ Lyman-$\alpha$ (Ly$\alpha$) emitters (LAEs) and covering 10 deg$^2$ in part of the SSA22 field with the Canada-France-Hawaii Telescope. We surveyed a comoving volume of $4.7\times 10^6$ Mpc$^3$ to a Ly$\alpha$ luminosity limit of $6.3\times10^{43}$ erg s$^{-1}$.
We look for Ly$\alpha$ candidates by applying the following criteria: i) clear emission line source, ii) no optical detections ($ugriz$ from CFHTLS), iii) no visible detection in the optical stack ($ugriz > 27$), iv) visually checked reliable NB$_J$ and $J$ detections and v) $J-K \leq 0$. We compute photometric redshifts and remove a significant amount of dusty lower redshift line-emitters at $z \sim 1.4 $ or $2.2$. A total of 13 Ly$\alpha$ candidates were found, of which two are marked as strong candidates, but the majority have very weak constraints on their SEDs.
Using follow-up observations with SINFONI/VLT we are able to exclude the most robust candidates as Ly$\alpha$ emitters. We put a strong constraint on the Ly$\alpha$ luminosity function at $z \sim 9$ and make realistic predictions for ongoing and future surveys. Our results show that surveys for the highest redshift LAEs are susceptible of multiple contaminations and that spectroscopic follow-up is absolutely necessary.
\end{abstract}


\section{Introduction}
Finding the first stars and galaxies is one of the most important tasks to test our understanding of galaxy formation in the early Universe. The current theoretical models of when and how these first galaxies were formed can only be tested and improved by reliable detections of galaxies at the highest redshifts. The confirmation of galaxies at a redshift of $z\sim 9-10$ would also allow the study of the epoch of reionization of the Universe. Measurements of the cosmic microwave background place this epoch at $z \sim 10.6$ \citep{Komatsu2011}, while \cite{Fan2006} located the end of the reionization epoch at a redshift of at least $z\sim 6$ by studying spectra of quasars at high redshift, where they found a lower limit to the neutral fraction of $\sim 10^{-3} - 10^{-2}$. 

 A widely used technique to detect very distant galaxies is the Lyman break technique (LBG), pioneered by \cite{Steidel1996} (see also \citealt{Guhathakurta1990}), which looks at a distinctive break in the UV spectrum of star-forming galaxies. More generally, one can use deep data in several broadbands to derive a redshift-probability distribution by fitting spectral energy distributions (SED) based on galaxy templates \citep[e.g.][]{McLure2011}.
 
Using the Lyman Break method, candidate galaxies have been found at very high redshifts \citep[$z \sim 7$, e.g.][]{Bouwens2011z7,Finkelstein2012,Oesch2012z8,McLure2012} and even $z\sim10$ \citep[]{Ellis2013,Oesch2013,Bouwens2013}, but the great majority of these are too faint to confirm spectroscopically. \cite{Lehnert2010} claimed the spectroscopic detection of a $z=8.6$ Lyman-$\alpha$ (Ly$\alpha$) line of a LBG in the Hubble Ultra Deep Field. However, \cite{Bunker2013} were unable to reproduce the detection with two independent sets of observations, leading to the suggestion that it could be an artefact. \cite{Brammer2013} found a tentative emission line that could be Ly$\alpha$ at $z = 12.12$ using the HST WFC3 grism, but this is only a $<3 \sigma$ detection and could be a lower redshift interloper. Recently, \cite{Finkelstein2013} report the detection of a Ly$\alpha$ emission line in a $z=7.51$ LBG, although the line is very close to a sky-line making identification significantly more difficult. Other attempts have been made, but so far no $z>7.5$ galaxy has been spectroscopically confirmed. There is a spectroscopic redshift determination of a $z=8.2$ Gamma Ray Burst \citep{Tanvir2009}, but not for its host.

Another successful technique to detect very high redshift ($z
\sim 4- 7$) galaxies is the narrow-band technique, which targets Lyman-$\alpha$ emitters (LAEs; e.g. \citealt{Pritchet1994,Thompson1994,Thompson1995,Hu1996,Cowie1998,Hu1998,Thommes1998,Rhoads2000,Rhoads2003,Rhoads2004,Fynbo2001,Hu2002,MalhotraRhoads2002,MalhotraRhoads2004,Fynbo2003,Ouchi2003,Hu2004,Taniguchi2005,Iye2006,Kashikawa2006,Shimasaku2006,Ouchi2008,Finkelstein2009,Ota2010,Hibon2011}).  Using the narrow-band technique one can search for sources with emission lines at specific redshifts, by looking at the excess the narrow-band has over the broadband. This way sources for which the continuum is too faint to be detected, can still be identified due to the bright emission lines. However, most emission line galaxies detected in narrow-band surveys are lower redshift interlopers such as H$\alpha$ and [O{\sc ii}] \cite[e.g.][]{Sobral2012}, which have to be identified using multiwavelength observations. Because the narrow-band is only sensitive to sources emitting in a small range of wavelengths, they can be used to look at a slice of redshifts and therefore a well-known comoving volume. Moreover, spectroscopic follow-up of high-redshift candidates is a priori easier for candidates detected by the narrow-band technique, as these candidates will have strong emission lines. Currently, the most distant spectroscopically confirmed NB-selected LAE is at a redshift of 6.96  \citep{Iye2006}, which is detected with narrow-band imaging from the Subaru telescope. 

\begin{figure}
\centering
\includegraphics[width=8cm]{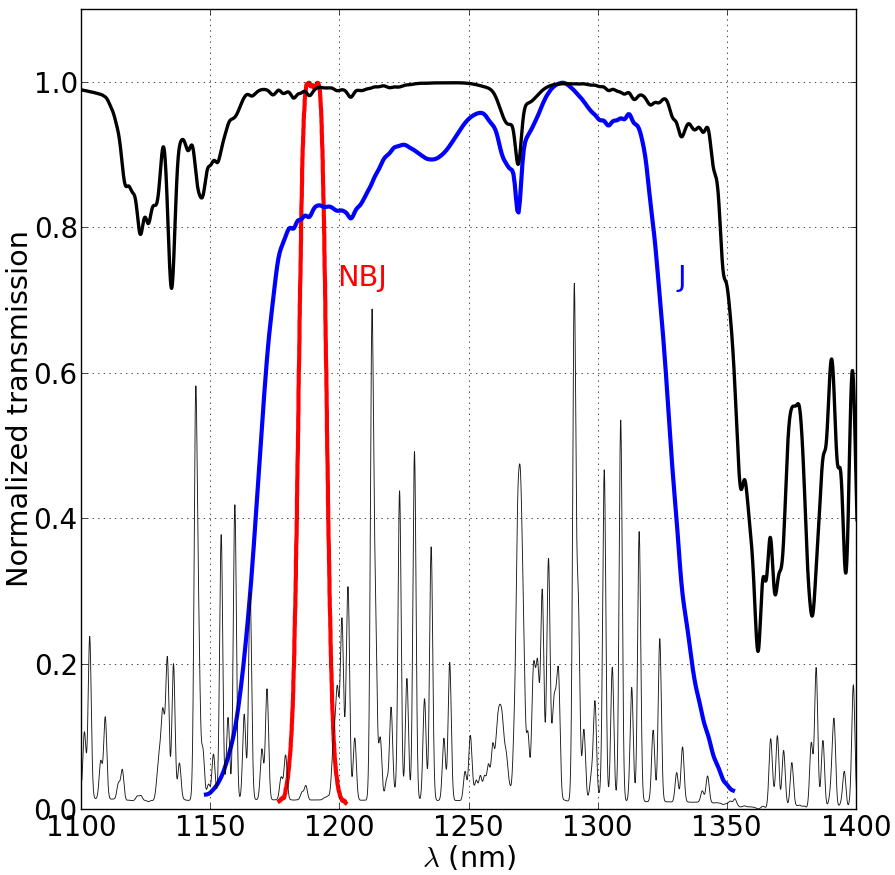}
\caption{\small{The atmospheric transmission in the near-infrared $J$ band, normalised to the maximum transmission for each curve. Atmospheric data is from Mauna Kea, Gemini Observatory \citep{Lord1992}, the airmass is 1.0 and water vapour column is 1.0 mm. The background (grey) atmospheric (OH) emission lines are also shown. The NB$_J$ filter used in this paper is transparent at wavelengths where there are no strong OH lines and also at wavelengths where the atmosphere is at its maximum transparency, thus allowing us to obtain deep observations in relatively little time.}}
\label{fig:filters}
\end{figure} 

To observe even higher redshift galaxies, observations in the near-infrared are required. Unfortunately, at these wavelengths there is significant foreground emission due to OH molecules in the Earth's atmosphere. Some OH windows exist at wavelengths where the atmosphere is transparant to radiation. It is possible to observe using narrow-band filters in these windows very effectively and several filters have been developed for this purpose (see Figure $\ref{fig:filters}$). Recent studies led to the identification of candidate Ly$\alpha$ emitters at a redshift of $z = 7.7$, but none of these has been spectroscopically confirmed yet \citep[]{Hibon2010,Tilvi2010,Clement2012,Krug2012,Jiang2013b}. 

Some attempts at somewhat higher redshifts ($z\sim9$) were made to detect Ly$\alpha$ \citep[]{WillisCourbin2005,Cuby2007,Willis2008,Sobral2009b}. The properties of such galaxies would provide strong tests of current models of galaxy formation and evolution and even the confirmation of just one luminous Ly$\alpha$ emitter at this redshift will be suitable for the study of these sources way before the next generation of telescopes, such as \emph{JWST} or the E-ELT. 

Ly$\alpha$ radiation is much more attenuated by a neutral intergalactic medium (IGM) than an ionized IGM, so large samples of Ly$\alpha$ emitters at these redshifts could be used to derive properties of the IGM at these early times. 

Current simulations \citep[e.g][]{Iliev2008} suggest that reionization started at the most overdense regions in the Universe, where ionizing sources nurtured expanding shells of ionized gas in the IGM. As Ly$\alpha$ radiation is easily absorbed by a neutral medium \citep{MalhotraRhoads2004}, Ly$\alpha$ emitters can only be observed once the ionized zone around them is large enough for the Ly$\alpha$ radiation to escape. This is expected to lead to a negative evolution in the Ly$\alpha$ luminosity function and dropping escape fraction of Ly$\alpha$ radiation at higher redshifts. Considerable effort has been put in spectroscopically studying the evolution of the Ly$\alpha$ line in Lyman break galaxies (LBGs) at high redshifts \citep[e.g.][]{Fontana2010,Pentericci2011,Vanzella2011,Ono2012,Schenker2012,Caruana2013,Finkelstein2013}. Recent non-confirmations and low success-rates at $z>7$ for their spectroscopic confirmation are interpreted as a signature that reionization is not yet completed at these redshifts. \cite{Treu2013}, for example, find that at $z\sim8$ Ly$\alpha$ emission of LBGs is suppressed by at least a factor of three.

For LAEs, it is found that up to at least a redshift of $z\sim6$ the Ly$\alpha$ luminosity function is remarkably constant \citep[e.g.][]{Shimasaku2006,Hu2004,Ouchi2008}. This indicates that LAEs are relatively more common and more luminous at earlier epochs, compared to LBGs (as the UV LF drops quickly in this redshift range; \citealt{Bouwens2007}). At $z\sim6-8$ there is evidence for evolution of the characteristic luminosity, but these samples, including failed attempts at $z=7.7$, can be significantly affected by cosmic variance, probing $\lsim 1$ deg$^2$ \citep[e.g.][]{Ouchi2010,Clement2012}.

At the bright end, however, the evolution could plausibly be very different. Luminous sources can ionize their own surroundings to allow Ly$\alpha$ photons to escape, as they redshift out of restframe-resonance wavelength in about 1 Mpc \citep[e.g.][]{Barton2004,CenHaiman2000,Curtis-Lake2012}. Furthermore, the observed clustering of Ly$\alpha$ emitters is expected to increase at higher redshift, as neighbouring sources will have larger overlapping ionized spheres and therefore a higher fraction of escaped Ly$\alpha$ photons \citep[e.g.][]{Ouchi2010}.

In order to find the most luminous Ly$\alpha$ emitters in the epoch of reionization which would be suitable for spectroscopic follow-up, we have undertaken the widest area search with a near infrared narrow-band filter to date. This paper is organised in the following way. \S2 presents the details of the observations, and describes the data reduction, calibrations and source extraction. \S3 presents the criteria for sources being selected as Ly$\alpha$ candidates and the results from the narrow-band search. \S4 presents the spectroscopic follow-up observations and results. \S5 discusses the results such as constraints on the Ly$\alpha$ $z=8.8$ luminosity function, and our survey is compared to past and future surveys. Finally, \S6 outlines the conclusions. A H$_0=70$\,km\,s$^{-1}$\,Mpc$^{-1}$, $\Omega_M=0.3$ and $\Omega_{\Lambda}=0.7$ cosmology is used  and all magnitudes are in the AB system, except if noted otherwise.

\section{NARROW-BAND OBSERVATIONS and DATA REDUCTION}
\begin{figure}
\includegraphics[width=8cm]{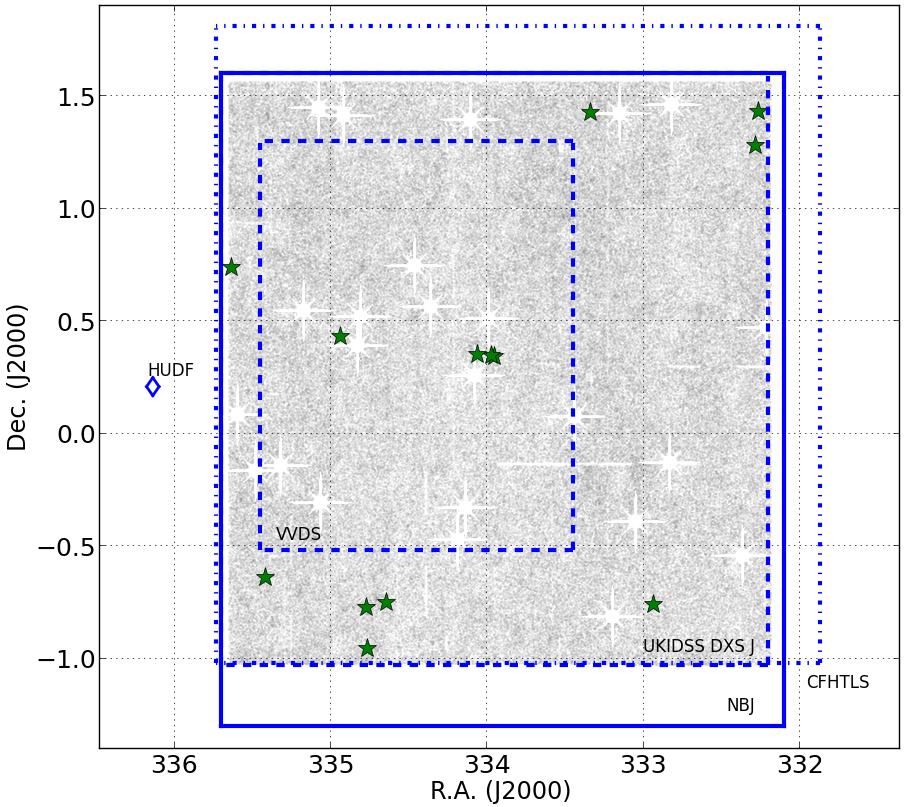}
\caption{\small{The surveyed area in the SSA22 field and overlap with other surveys. In grey we show all detected NB$_J$ sources, where white stars indicate the positions of the brightest stars ($J < 10.5$). NB$_J$ represents the area of the survey presented here. For Ly$\alpha$ emitters at $z = 8.76$, the surveyed area roughly corresponds to $\sim40\times60$ Mpc, with a depth of $\sim 180$ Mpc comoving. Our Ly$\alpha$ candidates are shown as green stars. The overlapping regions with CFHTLS W4 ($ugriz$), UKIDSS DXS ($JK$) \citep{Lawrence2007} and VVDS (spectro-$z$) \citep{Lefevre2005} are shown. For comparison, we also plot the size of the Hubble Ultra Deep Field, which is $\sim 3000$ times smaller than the area of this survey.}}
\label{fig:surveystrategy}
\end{figure}

During September-December 2011 and October-November 2012, we obtained medium depth narrow-band J photometry (NB$_J = 22.2$, $5\sigma$, F$_{lim}$ = $7 \times 10^{-17}$ erg s$^{-1}$ cm$^{-2}$) over a 10 deg$^2$ area in the SSA22 field using CFHT's WIRCam \citep{Puget2004} with a typical seeing of $0.6''$. The SSA22 field is the widest contiguous field for which a wealth of multi-wavelength data is available, most importantly $ugriz$ from CFHTLS-Wide and $JK$ from UKIDSS-DXS, see Fig. $\ref{fig:surveystrategy}$.

We use the LowOH2 filter ($\lambda_c = 1.187 \mu$m, $\Delta\lambda = 0.01\mu$m) which can detect Ly$\alpha$ emission ($\lambda_0 = 121.6$nm) at $z = 8.76 \pm 0.04$ in a comoving volume of $4.7\times 10^6$ Mpc$^3$. This is larger by at least half an order of magnitude compared to the largest previous survey. Detailed information on the observations, data reduction and general selection of emitters can be found in Sobral et al. (in prep.), but see also \cite{Sobral2013KMOS}. In this paper we explore potential Ly$\alpha$ candidates in the sample of emitters.

%
%

\subsection{Source Extraction and Survey Limits}
\label{extraction}
We use SExtractor \citep{Bertin1996} and detect $\sim 350,000$ sources across the 10 deg$^2$ narrow-band coverage. The 5$\sigma$ AB-magnitude limit for the survey is NB$_J = 22.2$, corresponding to an emission line flux limit of $7\times10^{-17}$ erg s$^{-1}$ cm$^{-2}$. This limit is computed by measuring the average background rms of the narrow-band images in empty 2$''$ diameter apertures, which is the aperture we use throughout the paper for all measurements. We note that because we use random aperture measurements, the rms that we measure already accounts for correlations in the noise. The limiting magnitude is converted to line-flux using the following formula:
\begin{equation} 
\label{eq:lineflux}
F_{\rm line} = \Delta\lambda_{NB_J}\frac{f_{NB_J}-f_J}{1-(\Delta\lambda_{NB_J}/\Delta\lambda_J)} 
\end{equation}
Here $F_{\rm line}$ is the line-flux (also called Ly$\alpha$ flux), $\Delta\lambda_{\rm NB_J}$ and $\Delta\lambda_{\rm J}$ ($\Delta\lambda_{\rm J} = 0.158\mu$m) are the widths of the narrow-band and broadband filter respectively, while $f_{\rm NB_J}$ and $f_{J}$ are the respective flux densities.

\section{Narrow-band selection of candidates}
In order to identify Ly$\alpha$ candidates, we look for line-emitters which show the characteristics of a $z>7$ source. These should have a Lyman-break, which should occur between the $z$ and $J$ band, and a flat or blue $J-K$ colour to exclude very dusty, lower redshift galaxies with strong breaks (e.g. the $4000 \,$\AA$ \,$ break). In practice, we use the following criteria:
\begin{enumerate}
\item Be selected as a line-emitter in Sobral et al. (in prep.) (as described in \S 3.1 below).
\item No detection in filters on the blue side of the $J$-band (see \S 3.2.1).
\item No visible detection in the stack of all optical bands (see \S 3.2.2).
\item Reliable excess between NB$_J$ and J (see \S 3.2.3).
\item $J-K \leq 0$ and a photometric redshift consistent with $z>4$ (see \S 3.2.4). 
\end{enumerate}

\subsection{Emission line candidates}
\begin{figure}
\centering
\includegraphics[width=8cm]{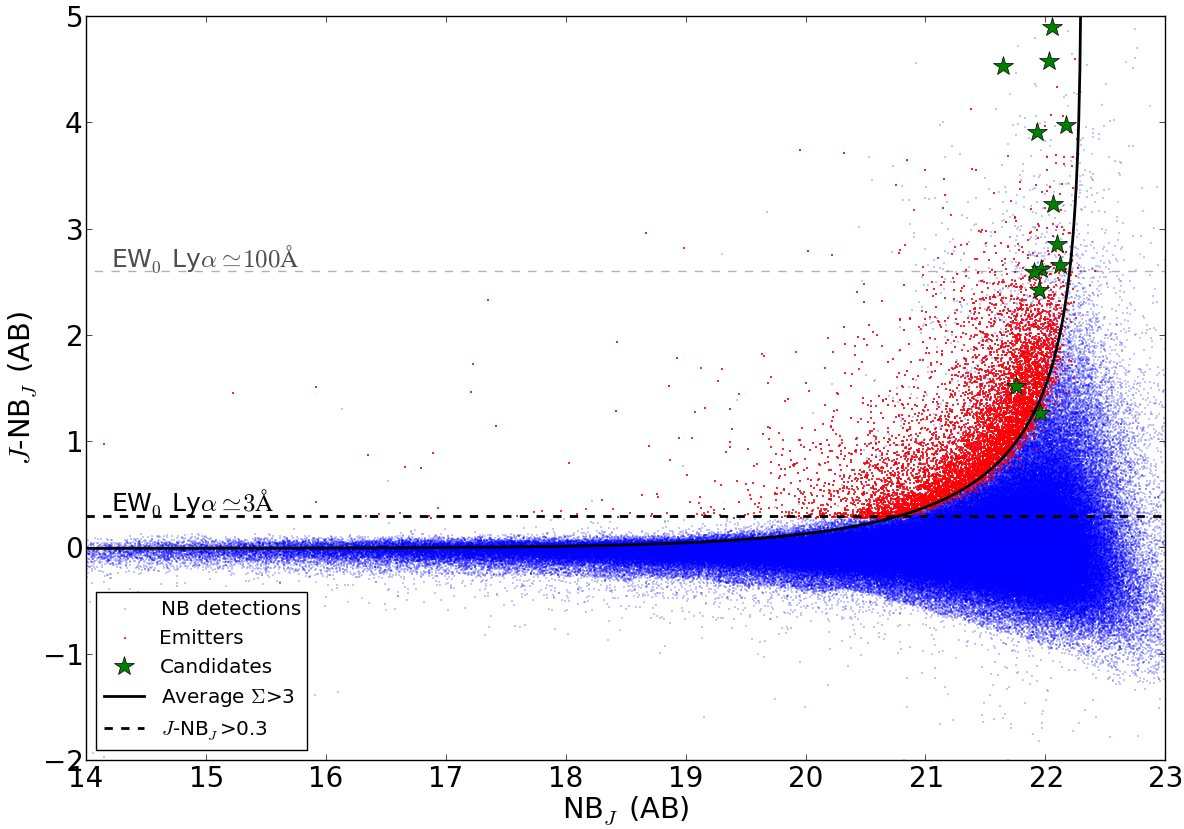}
\caption{\small{Colour-magnitude diagram for the NB$_J$ sources. The $J-$NB$_J$ colour is corrected using the $z$-band to compensate for the fact that the NB$_J$ filter is not in the center of the broadband, see Sobral et al. (in prep.) for more details. The dotted horizontal line is for an observed EW of 30 \AA, which corresponds to $J-$NB$_J$ $>$ 0.3. The $\Sigma = 3$ curve is shown for the average depth of the survey. Emitters are shown in red, as they have Equivalent Widths $> 30 $ \AA $\,$ and have a $\Sigma > 3$. The final Ly$\alpha$ candidates are shown with a green star and show a typical rest-frame EW (EW$_0$) of $\sim 100 $ \AA. The full sample of emitters is presented in Sobral et al. (in prep.).}}
\label{fig:jnbj}
\end{figure}

Emitters were selected using two criteria which quantify the excess the narrow-band has over the broadband. Firstly, the observed EW should be larger than 30\,\AA, corresponding to a rest-frame Ly$\alpha$ EW of $3\,$\AA . Secondly, the $\Sigma$ parameter (Eq. 2), which quantifies the significance of the narrow-band excess compared to the noise \citep{Bunker1995}, should be larger than 3 (similar to \cite{Sobral2013}). 
\begin{equation}
\Sigma=\frac{1-10^{-0.4(J-NB_J)}}{10^{-0.4(ZP-NB_J)}\sqrt{\pi r^2_{ap} (\sigma^2_{NB_J}+\sigma^2_{J})}}
\end{equation}
Where ZP is the zeropoint of the photometry (25), $r_{ap}$ is the radius of the apertures in pixels and $\sigma$ the RMS per pixel in each band. In case of non-detections in $J$, the detection limit was assigned. More detailed information of the procedure and the full sample of emitters will be presented in Sobral et al. (in prep). 
Using these criteria, out of the $\sim 350,000$ NB$_J$ sources individually detected, 6315 emitters were selected (see Fig. $\ref{fig:jnbj}$). This is after removing 2285 spurious sources and artefacts from bright stars by visual checks.

\subsection{Selecting Ly$\alpha$ candidates at $z=8.8$}
\subsubsection{Excluding lower redshift interlopers: optical broadband photometry}
A $z\sim$ 9 source should be undetected in filters on the blue side of the $J$-band, because the light at these wavelengths is absorbed by the IGM. This means that candidates must be undetected in the $u$, $g$, $r$, $i$ and $z$ bands. Data in these broadbands is available from the CFHT Legacy Survey (CFHTLS)\footnote{http://www.cfht.hawaii.edu/Science/CFHTLS/}. Deep data in the $J$ and $K$ bands is available from UKIDSS-DXS-DR10\footnote{http://www.ukidss.org/} ($J_{\rm AB}\sim23.4$, limit measured by the artificial star test). Two catalogues with sources in the optical bands of the CFHTLS were used. The first catalogue was the public CFHTLS-T0007 catalogue, in which sources were detected in the $gri$-stack. The second catalogue (Kim et al. in prep) contains 859,774 sources with photometric redshifts. It used $J$-band images from UKIDSS-DXS-DR10 for the detection on images. This catalogue is called the SSA22 catalogue and has depths of ($u$,$g$,$r$,$i$,$z$,$J$,$K$) = (25.2, 25.5, 25.0, 24.8, 23.9, 23.4, 22.9). For the optical these depths are taken from the public CFHTLS catalogue and correspond to 80\% completeness, for $JK$ these are 90\% completeness (Kim et al. in prep).

The line-emitters were matched to the CFHTLS and SSA22 catalogues with a maximum 1$''$ separation on the sky using {\sc topcat} \citep{TOPCAT}. A list with candidates that followed the first criterion was made by clearing sources with magnitudes brighter than the limits in one or more of the optical bands. After this first criterion, 302 candidates remained.
\begin{table}
\begin{center}
\label{tab:candidateselection}
\begin{tabular}{lr}
\hline
Step & Number \\
\hline
Line-emitters & 6315 \\
No optical detection & 302 \\
No detection in optical stack & 40 \\
Believable excess, NB$_J$, $J$ detections & 25 \\
Max number of Ly$\alpha$ candidates & 13 \\
With robust constraints & 2 \\ \hline
Fraction of H$\beta$/[O{\sc iii}] & 0.36 \\
Fraction of [O{\sc ii}] & 0.23 \\
Fraction of $z\sim3-6$ emission lines & 0.13 \\
Fraction of $z<0.8$ emission lines & 0.18 \\
Fraction of Ly$\alpha$ candidates & 0.10 \\
\hline
\end{tabular}
\caption{\small{Number of candidate Ly$\alpha$ emitters at $z = 8.8$ after each step and fractions of lower redshift interlopers out of the 302 sources without optical detection.}}
\end{center}
\end{table} 
\subsubsection{Visual check: optical stack}
For line-emitters that passed the first criterion, thumbnails were made of the stack of the optical bands $ugriz$. This is necessary to reject sources which have flux in the optical which is too faint to be detected in a single band, but that will be revealed in the stack as it has an estimated depth of $\sim27$ AB. Using the stack, sources with a detection in the optical (on the blue side of $J$) were identified and ruled out as $z = 8.8$ LAE. After this step, 40 candidates remained. Most of the candidates which were lost in this step are lower redshift contaminants such as [O{\sc ii}] at $z=2.2$, see Sobral et al. (in prep.). This is confirmed by their very red $J-K$ colours.

\subsubsection{Visual check narrow-band, broadband and excess}
Thumbnails are also made from the UKIRT $J$ and $K$ images and of the narrow-band image itself (see Fig A.1 and A.2 in the appendix and e.g. Fig. $\ref{fig:thumbs}$). Sources are then visually checked again in all bands. By comparing the broadband and narrow-band image, we were able to confirm if the source demonstrates a true narrow-band excess, instead of an excess caused by a boosted background. We also check whether the narrow-band flux density is consistent with that of the broadband $J$, because the broadband includes the narrow-band wavelength coverage. After all these visual checks 25 candidates remained, as 15 were marked as spurious or unreliable.

\begin{figure*}
\centering
\includegraphics[width=16cm]{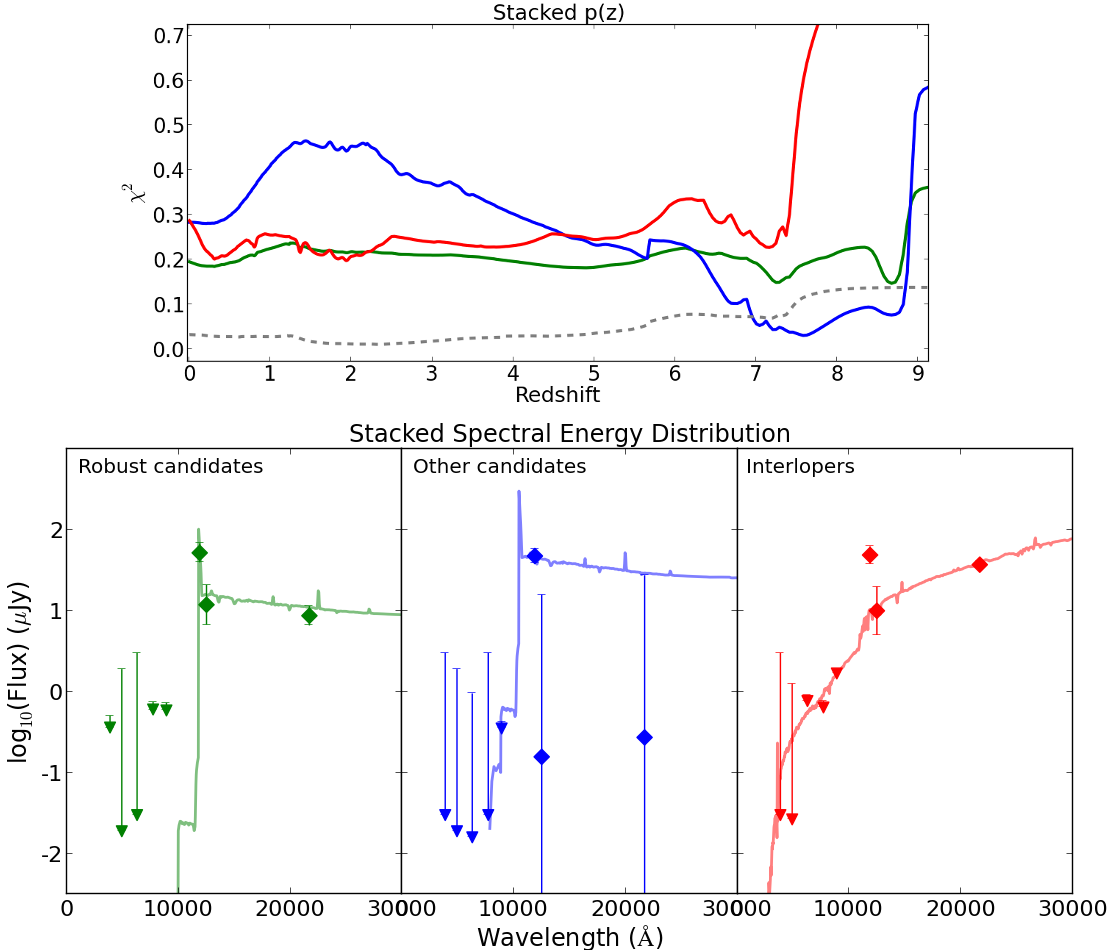}
\caption{\small{Top panel: Stacked redshift-$\chi^2$ distribution for the three samples. Bottom panel: Stacked Spectral Energy Distribution of: (left, green) our robust Ly$\alpha$ candidates ($z_{\rm phot}=8.7$), (centre, blue) the other Ly$\alpha$ candidates ($z_{\rm phot}=7.2$) and (right, red) the dominant lower redshift interlopers ($z_{\rm phot}=2.1$). For the interlopers the fits clearly prefer a dusty, red galaxy solution. In the top panel, dashed grey shows the redshift-$\chi^2$ distribution of the most robust candidates for running EAZY without adding the Ly$\alpha$ flux. The degeneracy between the [O{\sc ii}] and high redshift solution can clearly be seen in all three subsets. For the Ly$\alpha$ candidates the high redshift solution is preferred. }}
\label{fig:sed}
\end{figure*}

\subsubsection{Photometric redshifts}\label{foto}
Self-consistent photometry for the candidates was made by running SExtractor in dual-image mode on the thumbnails, using the narrow-band image as the detection image. In the case of non-detections by SExtractor in any of the other bands, the limiting magnitudes of the catalogue (see \S 3.2.1) were assigned. Using this consistent set of fluxes of the candidates in different wavelengths, we were able to derive a photometric redshift using EAZY\footnote{http://www.astro.yale.edu/eazy/}\citep{EAZY}. Unfortunately EAZY doesn't have a template for strong Ly$\alpha$ emission, therefore we create supplementary templates where we added this emission line to existing templates.

Some candidates at this point show a red $J-K$ colour and potentially very faint detections (below the 1-$\sigma$ limit) in the $r$, $i$ or $z$ band and are also not visible in the optical stack, indicating that these sources are likely very dusty lower redshift line-emitters. The emission line detected is in this case likely [O{\sc ii}] at $z = 2.2$ and the break between $z$ and $J$ the 4000 \AA \, break, which can mimic the Lyman break. From the 25 candidates for which we obtained an SED, 12 were marked as lower redshift contaminants. This left 13 candidates, which couldn't be further rejected without follow-up observations. We divide these candidates in different groups below.

\subsubsection{Different types of candidates}
\begin{figure}
\includegraphics[width=8cm]{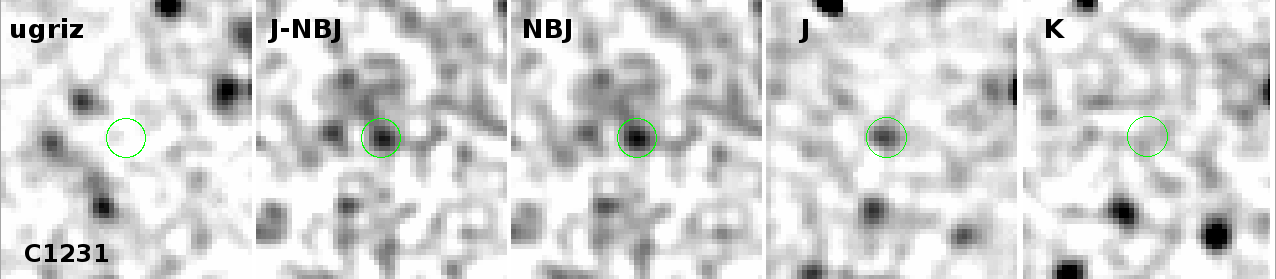}
\vspace{1.5ex}
\includegraphics[width=8cm]{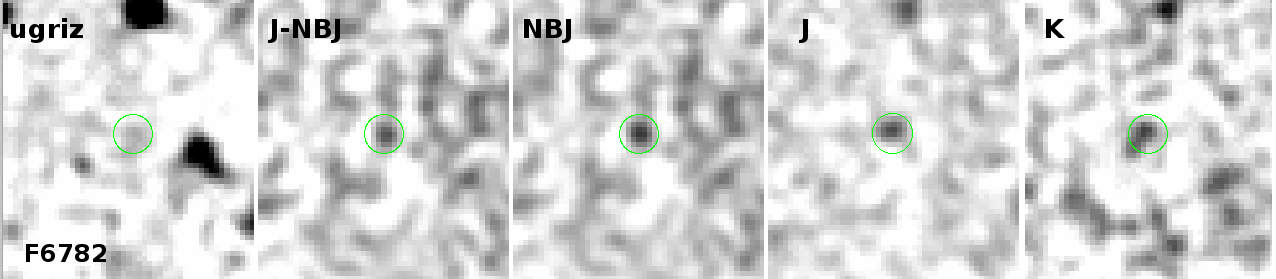}
\caption{\small{Thumbnail images for the most robust Ly$\alpha$ candidates. IDs C1231 and F6782. The sizes of the thumbnails are $15''\times15''$. The image on the left shows the stack of all the optical bands ($ugriz$), clearly there are no detections. The second image shows the excess, which consists of the difference between $J$ and NB$_J$. The right image shows the $K$-band. This is mainly used to check whether the source is not a very dusty emitter at a lower redshift, for example H$\beta$/[O{\sc iii}] at $z\sim$ 1.4 or [O{\sc ii}] at $z\sim$ 2.2, as such a source would be expected to have $J - K > 0$.
}}
\label{fig:thumbs}
\end{figure}

The candidates can be ordered in three different groups: i) candidates with detections in NB$_J$, $J$ and $K$, ii) candidates with NB$_J$ and $J$ detections and iii) candidates with only strong NB$_J$ detections. The measured magnitudes and computed quantities for individual candidates can be seen in Table A.1 in the appendix, which also shows how the candidates are grouped. The first group contains the two most robust sources with detections in $J$ ($>5\sigma$), best constrained $iz-J$ break, robust blue $J-K$ colours and best constrained SED, see Fig. $\ref{fig:thumbs}$ and Fig. $\ref{fig:sed}$. The second group consists of three candidates with both NB$_J$ and $J$, while the third group consists of 10 possible candidates with weak SED constraints and fainter $JK$ detections (see Fig. $\ref{fig:sed}$). Thumbnails for all candidates are shown in Fig. A.1 and Fig. A.2.


\subsubsection{Statistical likelihood}
In order to further investigate our selection, we stacked the thumbnails in all bands for the two robust candidates with best constrained broadband photometry, the 11 other candidates and the dominant lower redshift interlopers with individual photometric redshift of $\sim 2$. We measured the stacks with the narrow-band image as detection image and ran EAZY to compute photometric redshifts. As can be seen in Fig. $\ref{fig:sed}$, red and dusty galaxy templates are favoured for our lower redshift interlopers. Fig. $\ref{fig:sed}$ also shows that the Ly$\alpha$ candidates are best fitted by the high redshift solution, even though they show the same degeneracy as the lower redshift interlopers. By adding the strong Ly$\alpha$ emission in the EAZY templates, solutions around $z = 8.8$ are preferred. Without the addition of the Ly$\alpha$ flux to the template spectra, the lowest $\chi^2$ solutions would lie around $z\sim 2$, which can still be seen in the redshift-$\chi^2$ distribution (Fig. $\ref{fig:sed}$, top panel).

\subsection{Completeness}
The procedure of selecting emission line galaxies leads to potentially missing galaxies which have weak emission lines. To get an idea of how this influences the selection, we follow the procedure in \cite{Sobral2012,Sobral2013}. We compute the completeness by using a sample of sources which are 1) not selected as line-emitters, but 2) are selected as high redshift galaxies ($z>3$, using photometric redshifts and $BzK$ colours \citep{Daddi2004}. This selection resulted in a sample of $\sim$20,000 sources in our field and mimics our selection of Ly$\alpha$ candidates very well.
The second step is to add line flux to these sources and re-apply the selection criteria for sources being line-emitters (EW $> 30$ \AA, $\Sigma > 3$). This is done for increasing line-flux, and the number of sources which are being selected as line-emitters for each additional line-flux is counted. The completeness is the ratio of sources selected as emitters to the numbers in the original sample. This resulted in a completeness of $\sim90$\% for the average Ly$\alpha$ line-flux of the candidates of $\sim 1\times10^{-16}$ erg s$^{-1}$ cm$^{-2}$ and of $\sim75$\% for our detection limit of $7 \times 10^{-17}$ erg s$^{-1}$ cm$^{-2}$.

\subsection{Lower redshift contaminants}
Because of the large number of candidates with narrow-band excess consistent with $z > 7$ (302) when relying just on broadband photometry, it is possible to quantify the fraction of sources selected as $z>7$ candidates this way which are actually lower redshift interlopers. Interlopers were identified by using the optical stack and photometric redshifts (significantly improved by $K$-detections). From the 302 sources, 13 were marked as Ly$\alpha$ candidates, while 15 were marked as unreliable/spurious. This left 274 interlopers, meaning that 90\% of high redshift candidates which are selected just by using optical bands are lower redshift contaminants. Using the photometric redshifts available (see Sobral et al. in prep) we find that from this 90\% contamination, 36\% are H$\beta$/[O{\sc iii}] at $z=1.4$, 23\% [O{\sc ii}] at $z=2.2$ and 13\% higher redshift emitters such as carbon or magnesium lines from AGN at $z\sim3-6$, while the remaining 18\% are likely very faint lower redshift sources like Pa$\gamma$ at $z = 0.09$ or He{\sc ii} at $z = 0.44$.

\subsection{Properties of candidates}
Based on the narrow-band imaging, we find that our candidates are very luminous, with a median Ly$\alpha$ luminosity of L$_{\rm Ly\alpha} \simeq 1.0\times10^{44}$ erg s$^{-1}$. Compared to lower redshift ($z \sim 3-6 $) LAEs, the AGN fraction at these luminosities would be expected to be 100\%, although limited by small number statistics \citep{Ouchi2008}. On the other hand, sources with the similar or higher luminosities have already been found at $z=6-7$, such as the $z = 6.6$ giant LAE \emph{Himiko} \citep{Ouchi2009,Ouchi2013}, a triple major merger, with a Ly$\alpha$ luminosity $3.9\times10^{43}$ erg s$^{-1}$. \cite{Mortlock2011} found a quasar at $z=7.085$ with a Ly$\alpha$ luminosity of $\sim 10^{45}$ erg s$^{-1}$, ten times brighter than our candidates.

The majority of candidates show high EW$_{\rm obs}$ of $\sim 1000$ \AA $\,$ (see Table A.1 in the appendix), which is comparable to lower redshift samples \citep[e.g.][]{Ouchi2010} and a strong $iz-J$ break (median $\gsim 2$). Because of the clear $J$ detection, the strongest candidates have the lowest EW, but highest $iz-J$. We measured the FWHM of point sources around the candidates and of the stack of the robust candidates. The FWHM of point sources is $0.7\pm0.1 ''$, while the stack has a FWHM of $0.9\pm0.1 ''$. Converting this angular scale to a physical scale at $z = 8.8$ gives a physical size of $\sim 4$ kpc, which is roughly a fourth of the giant $z= 6.6$ LAE \citep{Ouchi2013} (which has extended Ly$\alpha$ emission), but a factor of four larger than 'typical' $z\sim 7-8$ LBG candidates \citep{Oeschsize}, so consistent with other observations.

All these properties estimated from the narrow-band imaging are physically realistic, but only spectroscopic follow-up can confirm the sources as real LAEs.

\section{Spectroscopic follow-up}
\subsection{Spectroscopic observations \& reduction}
\begin{figure}
\includegraphics[width=8cm]{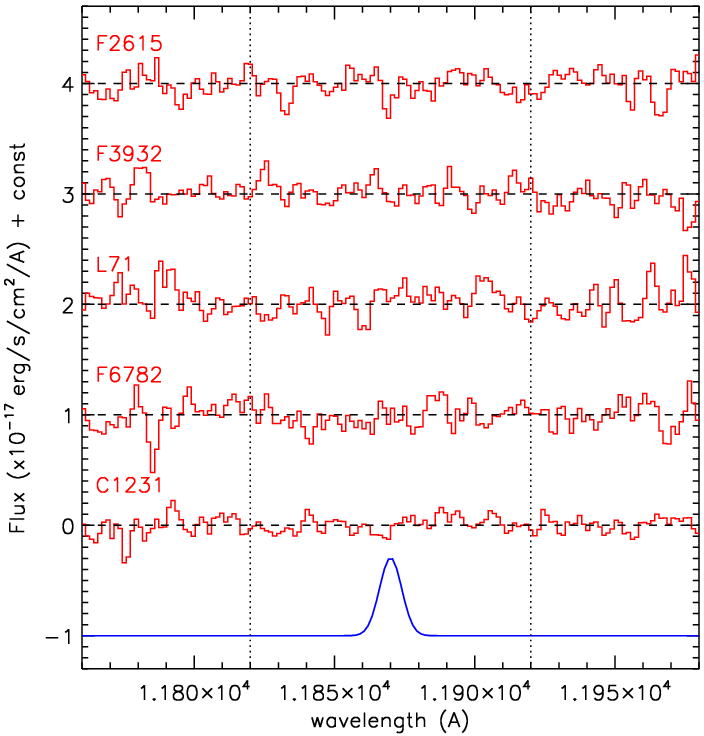}
\caption{\small{SINFONI IFU spectra from the five observed candidates (red). For illustrative reasons a constant is added to all fluxes except C1231. The bottom row (blue) shows how our emission lines should have looked based on the narrow-band estimated flux. The dashed vertical lines represent the width of the narrow-band filter.}}
\label{fig:spectro}
\end{figure}

SINFONI \citep{Eisenhauer2003,Bonnet2004} IFU observations of five Ly$\alpha$ candidates (C1231, F6782, F2615, F3932, L71, i.e. the two most robust and the three ones with highest significance, highest EW and lowest local noise properties) were taken as part of program 092.A-0786(A) between 2013 October 9 and 2013 October 31 in $< 0.8''$ seeing and photometric conditions.  Observations were made with the 8\,$\times\,8''$ field of view and the $J$ grating which has a resolving power of $\lambda$\,/\,$\Delta\lambda$\,=\,4000.  Each observation was split into eight 300 second exposures, and nodded around the target galaxy by $\sim 2 ''$ for sky subtraction purposes. Target C1231 was observed for a total of 4.8\,ks whilst the remaining four targets were observed for a total of 2.4\,ks. To reduce the data, we used the {\sc SINFONI} pipeline which extracts the slices, wavelength calibrates, flat-fields and sky-subtracts the data.  Additional sky subtraction was carried out using the techniques described in \cite{Davies2007}.  Flux calibration for each observation was carried out using standard star observations which were taken immediately before or after the science frames.  To search for line emission from the Ly$\alpha$ candidates, we extract a one-dimensional spectrum from the datacube, collapsed over a region with diameter of 1.2$''$ centered at the position of the narrow-band source, and show these in Fig. $\ref{fig:spectro}$.  These spectra have a noise of $0.7-1.1 \, \times 10^{-18}$\,erg\,s$^{-1}$\,cm$^{-2}$\,\AA$^{-1}$ over the wavelength range $1.182-1.192$\,$\mu$m (the approximate range of the narrow-band filter) and so a 3$\sigma$ detection limit for a line of width FWHM\,=\,250\,km\,s$^{-1}$ (typical for $z\sim7$ LAEs of \citealt{Ouchi2010}) of 1\,$\times$\,10$^{-17}$\,erg\,s$^{-1}$\,cm$^{-2}$. 
 As Fig. $\ref{fig:spectro}$ shows, none of the Ly$\alpha$ candidates are detected in emission by SINFONI, despite the flux limit of our narrow-band survey which should have yield $> 7 \sigma$ detections in all cases had emission lines been present. We also search for emission lines in the central $5\times5''$ coverage, but find nothing above $2\sigma$. We must conclude that, although the two most robust candidates can still be real Lyman-break galaxies based on their broadband magnitudes, they are excluded as luminous Ly$\alpha$ emitters at $z=8.8$. The others observed candidates are excluded as well. As these are the ones that resemble other candidates in the literature, their nature needs to be investigated (see \S 4.2). This has significant implications for other surveys.

One thing to note is that lower redshift line-emitters drawn from the same sub-sample (with similar excess significance and estimated line fluxes) were followed up with KMOS \citep{Sobral2013KMOS} and that strong emission lines were found in all of them.

\subsection{Contaminations to high-$z$ narrow-band searches: spurious sources, variability \& equatorial objects}
This section gives an explanation for none of the candidates being confirmed. To do this we again look at the different groups of candidates. 

i) We observed both candidates in the first group with the strongest NB$_J$, $J$ and $K$ photometry. The most likely explanation, given the relatively low $\Sigma$, but robust $J$ and $K$ (and given that the observations span different times), is that the excess is being boosted by noise. To estimate this, we look at the number of sources which are not selected as line-emitters, but fulfil the criteria of having no $ugriz$ and a blue $J-K$ colour and have reliable $J$ and $K$ detections, just as the two robust candidates in this group. From this number (306) we can compute that when looking at 3$\Sigma$ excess sources, we can expect 0.41 of these to have an excess by chance. The probability of getting both a 3.72 $\Sigma$ and a 3.03 $\Sigma$ source amongst the 306 is 1.1\%, this is low, but still possible.

ii) We observed the most robust candidate with only NB$_J$ and $J$ detections, and argue that, next to the possibility of the sources also being a statistical fluke, these sources are prone to variability. The time difference between the observations in $J$ and NB$_J$ is of order $1-2$ years. Because candidates are selected as having a narrow-band excess, variable sources which appear to be more luminous at the time when the narrow-band observations are taken than at the time when the broad-band observations are taken, lead to a false narrow-band excess. A rough estimate of variability is made by counting the number of sources with a very significant negative excess ($\Sigma <$ -7 and EW$<$-40\,\AA,) and excluding stars. We investigate whether any of these negative excess sources (300 in total) is caused by variability. By careful visual inspection of these sources (to determine whether the negative excess is real) we conclude that a fraction of 81 \% of these negative line emitters is a variable source. The other negative excess sources are binary stars or extended objects selected as two different sources in one of the filters by SExtractor. So in total a fraction of $7\times 10^{-4}$ ($0.81 \times \frac{300}{350000}$) of the line emitters is a variable source. This means that we can expect 4.4 line-emitters to be variable, possibly explaining the non-detection of our 3 Ly$\alpha$ candidates of this type.

iii) The candidates which only rely on a narrow-band detection have the chance of being a random noise spike, especially given that we observed a very wide area. We can get an estimate of the number of spurious sources in our survey by computing the total number of independent PSFs across the whole field. With a median seeing of $0.6''$ (Sobral et al. in prep) and an effective area of 9 deg$^2$, we have $3.2 \times 10^8$ PSFs. We computed local noise estimates around the candidates by taking the standard deviation from the counts in 1,000,000 2$''$ diameter apertures randomly distributed in $\sim 1.7$ arcmin$^2$ around the candidates, masking stars and other bright objects (NB$_J < 20$), see Table A1 in the Appendix. For the candidates in the third group, their median $\sigma$-detection is 5.44, based on the local noise. Using the number of PSFs, a total number of 8.5 spurious noise spikes is expected at this significance, which can explain the spectroscopic non-detection of the 8 candidates in this group. We have done a visual analysis to remove clearly spurious sources, such as those near stars or in noisy regions, but this analysis might have missed these random noise peaks. Also, as the SSA22 field is equatorial, there is a slight chance that we observe small solar system objects in our narrow-band and this could also contaminate searches in other equatorial fields.

\section{The LYA  Luminosity Function} \label{lfalpha}
\subsection{Volume corrections}
By assuming a top-hat filter profile, the comoving volume is $4.7\times 10^6$ Mpc$^3$, as our survey covered 9.0 deg$^2$, which is the area where the 10 deg$^2$ NB$_J$ survey overlaps with both the UKIDSS $J$ and CFHTLS $ugriz$ surveys.
The comoving volume must be corrected by including the dependency of the comoving volume on the luminosity, caused by the filter not being a perfect top-hat \citep[e.g][]{Sobral2009b,Sobral2013}. Making use of the derived luminosity limit of the narrow-band survey (namely $6.3\times10^{43}$ erg s$^{-1}$), it is possible to find the minimum luminosity for a source at a given redshift to be observed in the survey. 
For example, a source with a luminosity of $3.4\times10^{44}$ erg s$^{-1}$ would be detected at redshifts between 8.723 and 8.816 due to the filter transmisson and giving a corresponding comoving survey volume of $6.06\times10^6$ Mpc$^3$.

\subsection{Computing the Luminosity Function}
\begin{figure*}
\centering
\includegraphics[width=16cm]{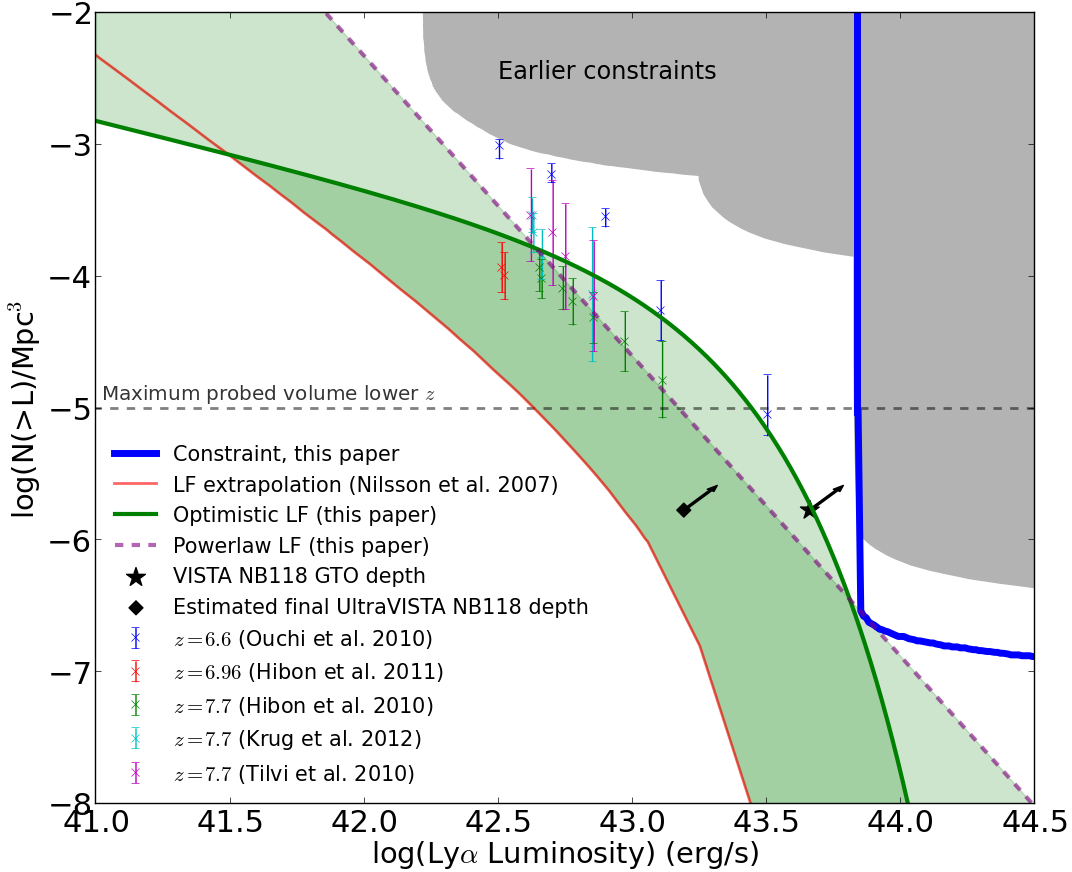}
\caption{\small{Constraint on the Ly$\alpha$ at z $\sim$ 9 luminosity function  of this paper compared to LFs at lower redshifts, a scaled LF extrapolation and optimistic fitted upper limit LF. The thick blue line shows the new constraint, drawn from the non-detections in our survey (after spectroscopic follow-up). The new constraint improves previous ones by a factor of five. The thick green line is an optimistic fitted Schechter function based on our observations and earlier observations at $z=7.7$, while the magenta line shows a fitted power law. The red line is an extrapolation from luminosity functions at lower redshift. The green area marks the region where we expect to observe LAEs, where there is a higher chance in the darker region. Also shown are the points from lower redshift narrow-band searches. We plot the point of the depth of the finished VISTA NB118 GTO survey \citep{Milvang-Jensen2013} and make a realistic estimate of what the depth will be of the ongoing UltraVista NB118 survey \citep{McCracken2012}.}}
\label{fig:lfcomparison}
\end{figure*}

Following the non-detections in our spectroscopic follow-up, we put a constraint on the bright end of the $z=8.8$ Ly$\alpha$ luminosity function by probing to a Ly$\alpha$ luminosity of $10^{43.8}$ erg s$^{-1}$ over $4.7\times10^6$ Mpc$^3$, (see Fig. $\ref{fig:lfcomparison}$).
Using literature data from $z = 7.7$ LAE searches (see Table 2) we compute an optimistic upper limit to the luminosity function, using all sources from earlier $z= 7.7$ surveys and our constraint as an upper-limit. Although $z=7.7 - 8.8$ seems a significant difference in redshift, the difference in cosmic time is comparable to a sample of e.g. $z=0.78-0.82$. Fitting a Schechter-function with a fixed faint-end slope $\alpha$ of $-1.5$ (following \citealt{Ouchi2010}), we find log$_{10}$($\Phi^*$)$\,= -4.21^{+0.11}_{-0.11}$ and log$_{10}$($L^*$)$\,= 43.10^{+0.03}_{-0.03}$. We also fit a simple power law with log$_{10}$($\Phi$) $= 93.3 -2.28$log$_{10}$($L$). We note that these LFs should be interpreted as a very optimistic scenario, as none of the $z=7.7$ sources have been confirmed spectroscopically.

\begin{table*}
\begin{center}
\caption{\small{Narrow-band Ly$\alpha$ surveys at z $>$ 7.}}
\begin{tabular}{lrrrrr}
\hline
Reference & Area  & Depth      & $z$ & No. LAE & Field \\
 & (arcmin$^2$) & ($10^{42}$ erg s$^{-1}$) &  & &   \\ \hline
  Ota et al. 2010 & 4680 & $9.2$  & 7 & 3& SXDS\\
  Tilvi et al. 2010 & 784 & $4$  & 7.7 & 4& LALA Cetus\\
  Hibon et al. 2010 & 400 & $6$  & 7.7 & 7& CFHT-LS D1\\
  Hibon et al. 2011 & 465 & $\sim1$  & 6.96 & 6& COSMOS\\
  Cl{\'e}ment et al. 2012 & 169 & $\sim2$  & 7.7 & 0& Bullet, GOODS-S, CFHT-LS D4\\
  Krug et al. 2012 & 760 & $5.5$  & 7.7 & 4& COSMOS\\ \hline
  Willis $\&$ Courbin 2005 & 6.25 & $20$  & $\sim$ 9 & 0& HDF South\\
  Willis et al. 2008 & 12 & $10$  & $\sim$ 9 & 0& Abell 1689, 1835, 114\\
 Cuby et al. 2007 & 31 & $13$  & 8.8 & 0& GOODS\\
  Sobral et al. 2009b & 5040 & $63$  & 8.96 & 0& COSMOS, UDS\\
\bf This paper & 32400 & $63$  & 8.76 & 0& SSA22 \\ \hline
\end{tabular}
\end{center}
\label{tab:othersurveys}
\end{table*}

\subsection{Comparison with other surveys}
Earlier searches for Ly$\alpha$ at $z \sim 9$ have put constraints on the luminosity function. \cite{Cuby2007} and \cite{Willis2008} got to faint magnitudes, but observed significantly smaller areas ($\sim 10$ arcmin$^2$). \cite{Sobral2009b} is of the same depth as our current survey, but probed a factor of five smaller area. More searches have been conducted at a redshift of $\sim 7.7$, which has not led to any spectroscopic confirmation, despite the recent attempts. For a summary, see Table 2. 

Recently \cite{Jiang2013b} followed-up one $z = 7.7$ candidate, but they failed to confirm the line. Compared to our survey, this candidate was a factor 10 fainter and detected in a probed volume 200 times smaller than this work. None of the other candidates from the $z=7.7$ searches have been confirmed spectroscopically so far. Faisst et al. (in prep) also followed-up two of the best \cite{Krug2012} $z=7.7$ candidates, finding no line emission, in line with our results. The majority of these $z= 7.7$ candidates rely on narrow-band detections only, as our group iii) candidates (\S 3.2.5) and we caution about these candidates being real, based on the arguments in \S 4.3.

\subsection{Predictions for future and on-going surveys}
Between $z \sim 6$ and $z \sim 7$, \cite{Ouchi2010} find negative evolution in the Ly$\alpha$ LF, caused mostly by a fainter $L^*$. It would be a logical step to further investigate this to even earlier times. In order to examine the evolution of the bright end of the LF, samples of luminous Ly$\alpha$ candidates would have to be compared with lower redshift samples. This is however impossible because the majority of searches only probes small survey areas (see Table 2). It is by no means certain that extrapolation \citep[e.g.][]{Nilsson2007EXTR} from results over smaller areas (and fainter LAEs) hold for large luminosities as those surveys miss the most luminous sources. As illustrated in Fig. $\ref{fig:lfcomparison}$ there is practically nothing known at densities log $\Phi < -5$ at $z\sim 6-8$, but also not for $z\sim5-6$ \citep{Ouchi2008}. LFs at $z<6$ on the bright-end are dominated by cosmic variance. The only way to overcome these problems is to conduct very wide ($\sim 10$ deg$^2$) narrow-band searches for LAE at redshifts $z = 2 - 8$, as for example Subaru's Hyper Suprime Cam will be able to do over the next few years \citep{Takada2012}.

Future surveys such as the NB118 ($\lambda_c = 1.19$ $\mu$m, $\Delta\lambda = 12.3 $nm) UltraVista  component \citep{McCracken2012} will probe deeper than this work and should have a good chance of finding Ly$\alpha$ emitters at $z \sim 9$, although their survey area will be smaller. Because of a smaller survey area the survey will be less likely to find the most luminous sources and will be more affected by cosmic variance, which plays an important role in whether high redshift LAEs can be observed due to reionization-topology \citep[e.g.][]{TaylorLidz2013}. The completed VISTA NB118 GTO survey has probed 0.9 deg$^2$ to a line flux of $4.4-5\times10^{-17}$ erg s$^{-1}$ cm$^{-2}$ (depending on reduction \citep{Milvang-Jensen2013}), which is already fainter than our current constraint (see Figure $\ref{fig:lfcomparison}$). The ongoing UltraVISTA NB118 survey will probe the same area in a different part of the sky and its depth can be extrapolated from the finished survey. As the VISTA NB118 GTO survey had 12\% (12.33 h/px) of the total exposure time of the ongoing UltraVISTA NB118 survey (112 h/px; \citealt{McCracken2012,Milvang-Jensen2013}), we estimate the final depth assuming that the depth increases with $\sim t_{exp}^{0.5}$. Therefore the final depth of the UltraVISTA NB118 survey will increase by a factor $\sim 2.9$, leading to a Ly$\alpha$ flux limit of $\sim 1.5-1.7\times10^{-17}$ erg s$^{-1}$ cm$^{-2}$.

Using our optimistic upper limit to the luminosity function and the \cite{Nilsson2007EXTR} extrapolation as lower limit, we can estimate the number of LAEs that the UltraVista survey will detect (also see the marked green area in Fig. $\ref{fig:lfcomparison}$). For the GTO depth, this number is $0.001 - 1.19$ and for the estimated final UltraVista depth this is $0.19-23.35$. When fitting a power law luminosity function to the $z= 7.7-8.8$ points, the current depth is expected to find $0.47$ LAE, while the final depth can get to $6.00$ LAEs. Since the UltraVista camera has 16 detectors with 16 different NB118 filters there are small variations on the sky brightness from detector to detector, so some regions are shallower and some deeper \citep{Milvang-Jensen2013}. We used the median depths for the estimates above, but the numbers could vary because of a higher contribution from the deepest regions. 
Nevertheless, it is possible that even with the finished UltraVISTA NB118 survey, no $z=8.8$ LAE will be found. On the other hand, our results clearly show how important it will be to spectroscopically follow-up any candidate arising from any similar survey.

\section{Conclusions}
We have conducted a very wide narrow-band survey over 10 deg$^2$ in the near infrared and identified 6315 line-emitters using a $1.19 \mu$m narrow-band filter. In this work we identified possible $z = 8.8$ Ly$\alpha$ candidates in the sample of line-emitters and followed-up the strongest ones spectroscopically. The main conclusions are: 

\begin{itemize}
\item A significant fraction ($\sim 300$) of the line-emitters are consistent with being at high redshift ($z > 3$), of which some might be Ly$\alpha$ at $z = 8.76$. This narrow-band survey increased the probed volume by half an order of magnitude compared to previous surveys and is thus sensitive to the rarest and most luminous sources. 
\item By doing careful visual checks of the robustness of the detections and by excluding line-emitters which are detected in any of the optical bands and which show a red $J-K$ colour, we find 13 possible Ly$\alpha$ candidates. We order them in different groups based on their broadband photometric constraints. The two most robust candidates have reliable detections in narrow-band, strongest constraints from photometric redshifts, $iz-J$ break and robust $J$ detections.
\item 90\% of the high redshift candidate line-emitters, selected on having no/very faint flux in the optical, are lower redshift interlopers. By including the $K$ band and computing photometric redshifts we find that approximately 40\% are H$\beta$/[O{\sc iii}] at $z=1.4$, 25\% [O{\sc ii}] at $z=2.2$ and 15\% are candidate AGN emission-lines (e.g. carbon or magnesium) at $z\sim3-6$, while the remaining are likely very faint lower redshift sources like Pa$\gamma$ at $z = 0.09$ or He{\sc ii} at $z = 0.44$.
\item Spectroscopic follow-up of the two most robust Ly$\alpha$ at $z=8.8$ candidates, two sources with the largest EWs and another with brightest J failed to confirm these sources as line-emitters. This is probably caused by a combination of spurious sources, variability and (although unlikely) solar system objects. This result has very strong implications to current and future candidates for LAEs at $z=7.7$ and $z>8$.
\item After the follow-up, we put the strongest constraints on the bright end of the luminosity function with half an order of magnitude improvement in the probed volume and it could still mean little to no evolution in the luminous end.
\item Using an optimistic upper-limit to the LF and a lower redshift extrapolation, we estimate the number of LAEs that will be detected by the completed VISTA NB118 GTO survey to be between $0.001$ and $1.19$ with the current depth and to be between $0.19$ and $23.35$ for the estimated final depth of the ongoing UltraVista NB118 survey.
\item Because of the lack of a comparably wide surveys, it is difficult to study the evolution of the bright end of the LF and extrapolations from other considerably smaller surveys at lower redshifts are unusable. Although the number density of Ly$\alpha$ emitters is expected to decline at higher redshifts, this isn't necessary the case for the bright end of the luminosity function, because of the topology of reionization. It is therefore of utmost importance to study the bright end of the Ly$\alpha$ luminosity function at lower redshifts in order to understand the evolution in the LF completely.
\item As our strongest candidates looked realistic in the images and had realistic physical properties based on the photometry, but still are not confirmed, we highlight the necessity for all other surveys to do this spectroscopic follow-up, especially when candidates are based on just a single-band detection. This has significant consequences for any similar and for deeper surveys, clearly pointing out that despite sources passing all tests, only spectroscopic observations can confirm them.
\end{itemize}

\section*{Acknowledgments}
We thank the anonymous referee for the comments and suggestions which improved both the quality and clarity of this work. DS acknowledges financial support from the Netherlands Organisation for Scientific research (NWO) through a Veni fellowship. IRS acknowledges support from STFC (ST/I001573/1), a Leverhulme  Fellowship, the ERC Advanced Investigator programme DUSTYGAL 321334 and a Royal Society/Wolfson Merit Award. PNB acknowledges support from the Leverhulme Trust. JWK acknowledges the support from the Creative Research Initiative program, No. 2008-0060544, of the National Research Foundation of Korea (NRF) funded by the Korea government (MSIP). JPUF and BMJ acknowledge support from the ERC-StG grant EGGS-278202. The Dark Cosmology Centre is funded by the Danish National Research Foundation.
This work is based in part on data obtained as part of the UKIRT Infrared Deep Sky Survey.
Based on observations obtained with MegaPrime/MegaCam, a joint project of CFHT and CEA/IRFU, at the Canada-France-Hawaii Telescope (CFHT) which is operated by the National Research Council (NRC) of Canada, the Institut National des Science de l'Univers of the Centre National de la Recherche Scientifique (CNRS) of France, and the University of Hawaii. This work is based in part on data products produced at Terapix available at the Canadian Astronomy Data Centre as part of the Canada-France-Hawaii Telescope Legacy Survey, a collaborative project of NRC and CNRS. 
This work was only possible due to OPTICON/FP7 and the access that it granted to the CFHT telescope.
The authors also wish to acknowledge the CFHTLS and UKIDSS surveys for their excellent legacy and complementary value - without such high quality data-sets this research would not be possible.

\bibliographystyle{mn2e.bst}
\bibliography{bibliographyJM}

\appendix
\input{Appendix}

\bsp

\label{lastpage}

\end{document}

%% file: Appendix.tex
\appendix

\section{Lyman-$\alpha$ candidate data}

\begin{landscape}
\begin{table}
\centering
\begin{tabular}{|l|r|r|r|r|r|r|r|r|r|r|r|r|r|r|r|r|c|r|}
  \multicolumn{1}{|c|}{ID} &
  \multicolumn{1}{c|}{R.A.} &
  \multicolumn{1}{c|}{Dec.} &
  \multicolumn{1}{c|}{$u$} &
  \multicolumn{1}{c|}{$g$} &
  \multicolumn{1}{c|}{$r$} &
  \multicolumn{1}{c|}{$i$} &
  \multicolumn{1}{c|}{$z$} &
  \multicolumn{1}{c|}{NB$_J$} &
  \multicolumn{1}{c|}{$\Delta$NB$_J$} &
  \multicolumn{1}{c|}{$J$} &
  \multicolumn{1}{c|}{$\Delta J$} &
  \multicolumn{1}{c|}{$K$} &
  \multicolumn{1}{c|}{$\Delta K$} &
  \multicolumn{1}{c|}{EW$_{\rm obs}$} &
  \multicolumn{1}{c|}{$\Sigma$} &
    \multicolumn{1}{c|}{$\sigma_{\rm NB,local}$} &
  \multicolumn{1}{c|}{L$_{\rm Ly\alpha}$} & 
  \multicolumn{1}{c|}{$z_{\rm phot}$}\\
  & (J2000) & (J2000) & & & & & & & & & & & &(\AA) &  & & ($10^{44}$ erg s$^{-1}$) &\\
\hline
  C1231* & 334.642 & $-0.751$ & $>25.2$ & $>25.5$ & $>25.0$ & $>24.8$ & $>23.9$  & 21.75 & 0.21 & 23.32 & 0.22 & 24.01 & 0.55 & 450 & 3.72 & 10.32 & 1.19 & 8.68$^{+0.05}_{-0.07}$\\
   F6782* & 333.335 & 1.426 & $>25.2$ & $>25.5$ & $>25.0$  & $>24.8$ & $>23.9$ & 21.95 & 0.23 & 23.25 & 0.21 & 23.22 & 0.23 & 219 & 3.03 & 8.09 & 0.90 & 8.66$^{+0.06}_{-1.87}$\\ \hline
   F2615* & 334.768 & $-0.775$ & $>25.2$ & $>25.5$ & $>25.0$ & $>24.8$  & $>23.9$  & 21.94 & 0.28 & 24.38 & 0.59 & $>22.9$ & $-$ & 401 & 3.45 & 4.30 & 1.18 & 0.73$^{+4.55}_{-0.42}$\\ 
  C1591 & 334.939 & 0.429 & $>25.2$ & $>25.5$ & $>25.0$  & $>24.8$ & $>23.9$ & 21.97 & 0.26 & 24.63 & 0.69 & $>22.9$ & $-$ & 604 & 3.65 & 6.16 & 1.18 & 6.81$^{+0.13}_{-0.56}$\\
  C83 & 332.282 & 1.28 & $>25.2$ & $>25.5$ & $>25.0$ & $>24.8$ & $>23.9$  & 21.90 & 0.24 & 24.54 & 0.64 & $>22.9$ & $-$ & 2108 & 3.86 & 9.94 &  1.25 & 6.99$^{+1.07}_{-0.13}$\\ \hline
  C803 & 333.949 & 0.341 & $>25.2$ & $>25.5$ & $>25.0$  & $>24.8$ & $>23.9$  & 22.06 & 0.26 & $>23.4$ & $-$ & $>22.9$  & $-$ & $>241$ & 3.61 & 4.92 & 1.12 & 6.92$^{+1.45}_{-0.36}$\\
  F1035 & 332.265 & 1.431 & $>25.2$ & $>25.5$ & $>25.0$ & $>24.8$  & $>23.9$ & 21.95 & 0.21 & $>23.4$ & $-$  & $>22.9$ & $-$ & $>286$ & 3.21 & 7.99 & 1.30 & 2.07$^{+1.23}_{-1.56}$\\
  F1024 & 335.415 & $-0.643$ & $>25.2$ & $>25.5$ & $>25.0$ & $>24.8$  & $>23.9$  & 22.27 & 0.31 & $>23.4$ & $-$ & $>22.9$& $-$ & $>165$ & 3.28 & 7.51 & 1.04 & 8.68$^{+0.03}_{-2.41}$\\
  F4815 & 333.968 & 0.345 & $>25.2$ & $>25.5$ & $>25.0$  & $>24.8$  & $>23.9$ & 22.09 & 0.31 & $>23.4$ & $-$ & $>22.9$& $-$ & $>231$ & 3.21 & 4.79 & 1.06 & 7.36$^{+0.34}_{-0.29}$\\
  F3932* & 335.633 & 0.737 & $>25.2$ & $>25.5$ & $>25.0$ & $>24.8$  & $>23.9$  & 21.61 & 0.17 & $>23.4$ & $-$ & $>22.9$& $-$ & $>403$ & 3.83 & 9.69 & 1.72 & 8.68$^{+0.06}_{-0.10}$\\
  F2751 & 334.766 & $-0.958$ & $>25.2$ & $>25.5$ & $>25.0$ & $>24.8$  & $>23.9$  & 22.12 & 0.29 & $>23.4$  & $-$ & $>22.9$ & $-$ & $>224$ & 3.11 & 5.96 & 1.03 & 1.62$^{+0.64}_{-1.02}$\\
  F4818 & 334.056 & 0.350 & $>25.2$ & $>25.5$ & $>25.0$  & $>24.8$ & $>23.9$ & 22.03 & 0.29 & $>23.4$ & $-$ & $>22.9$ & $-$ & $>253$ & 3.64 & 4.86 & 1.21 & 7.28$^{+0.58}_{-0.89}$\\
  L71* & 332.934 & $-0.761$ & $>25.2$ & $>25.5$ & $>25.0$ & $>24.8$  & $>23.9$ & 22.06 & 0.30 & $>23.4$ & $-$ & $>22.9$ & $-$ & $>243$ & 3.42 & 4.88 & 1.18 & 1.94$^{+1.91}_{-1.41}$\\
\hline\end{tabular}

\caption{List of Lyman-$\alpha$ candidates. Candidates observed with SINFONI are marked with a *. The first two are the candidates which are marked as being most robust. The next three also seem to have $J$ detections, while the last group relies mostly on their NB$_J$ detection. The magnitudes are in AB and estimated with the narrow-band as detection image. All optical measurements are on the noise level, as are some $J$ and $K$ measurements. The limit is assigned for these non-detections. In the case of believable detections, we measured NB$_J$, $J$ and $K$ in single mode as this is less prone to astrometric errors. $\Delta$NB$_J$, $\Delta J$ and $\Delta K$ are the 1$sigma$ magnitude errors by SExtractor. EW$_{\rm obs}$ is in \AA\, and a lower limit is assigned for non-detection in $J$. The excess-significance $\Sigma$ is estimated from single mode photometry of the NB$_J$ and $J$ band, with a local lower limit for the non-detections. $\sigma_{NB,local}$ is the significance level in 2$''$ diameter aperture measurements from the local ($\sim 1.7$ arcmin$^2$) area around the candidates. Luminosities are based on a redshift of $z=8.76$. The photometric redshifts are calculated with EAZY. $z_{\rm phot}$ includes the narrow-band filter and a artificial Lyman-$\alpha$ template, 1$\sigma$ errors are shown. It should be noted however that these are largely unconstrained, especially for the candidates in the third group.}

\end{table}
\end{landscape}
\begin{figure*}
\centering
\includegraphics[width=14cm]{./THUMBS/C1231.png}
\vspace{0.5ex}
\includegraphics[width=14cm]{./THUMBS/F6782.png}
\vspace{0.5ex}
\includegraphics[width=14cm]{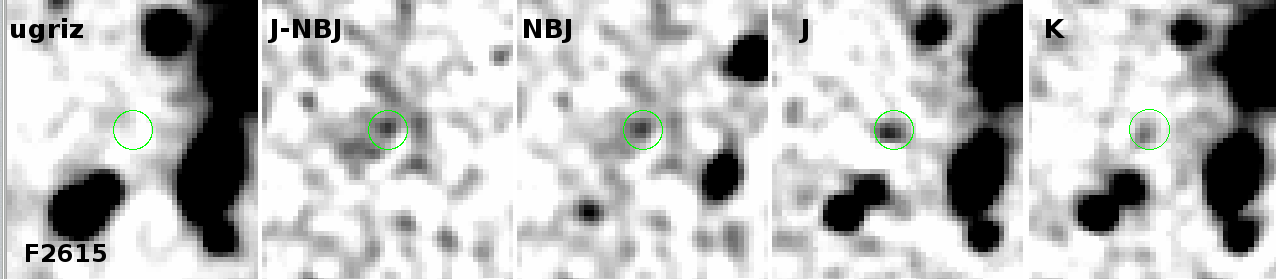}
\vspace{0.5ex}
\includegraphics[width=14cm]{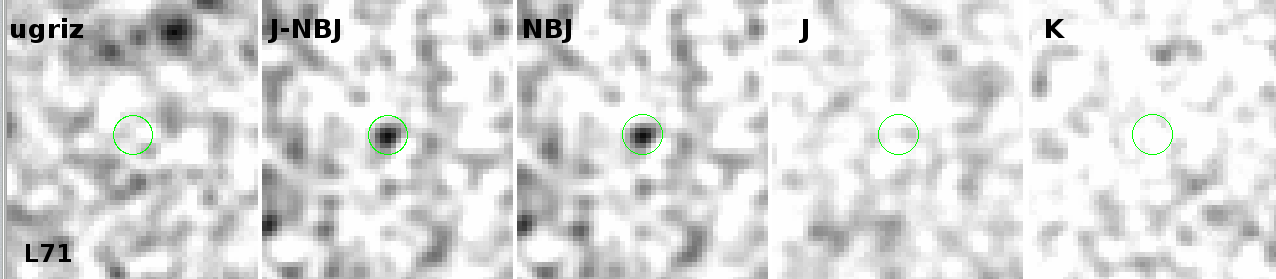}
\vspace{0.5ex}
\includegraphics[width=14cm]{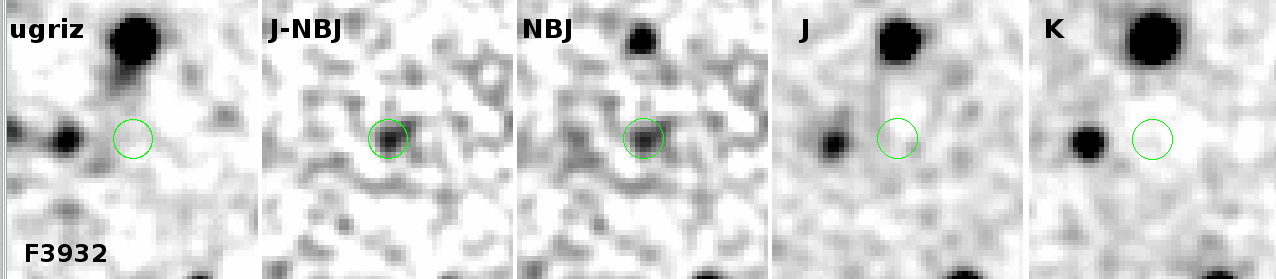}
\vspace{0.5ex}
\includegraphics[width=14cm]{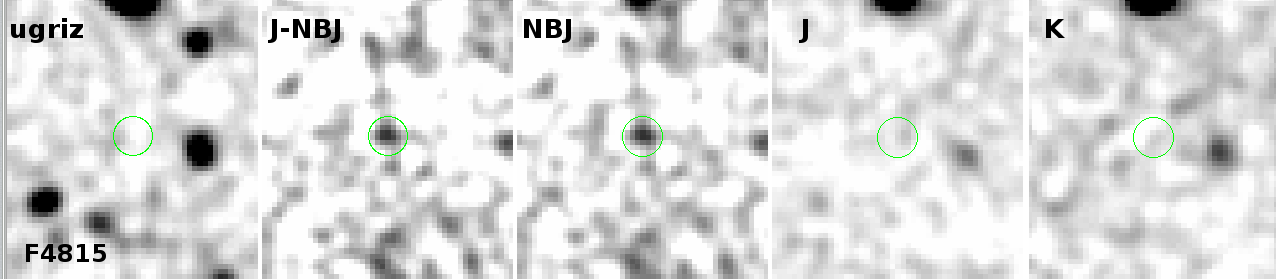}
\caption{\small{Top: Thumbnails for the five Lyman-$\alpha$ candidates which were followed-up spectroscopically. Bottom: one of the other candidates. Circles are placed at the center position of the thumb, corresponding to the position of the detection in the narrow-band. The angular scale of the thumbnails is $15\times15 ''$.}}
\end{figure*} 

\begin{figure*}
\centering
\includegraphics[width=14cm]{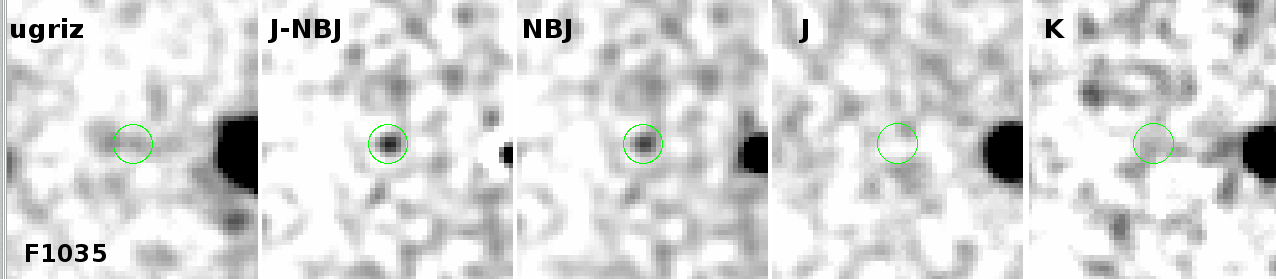}
\vspace{0.5ex}
\includegraphics[width=14cm]{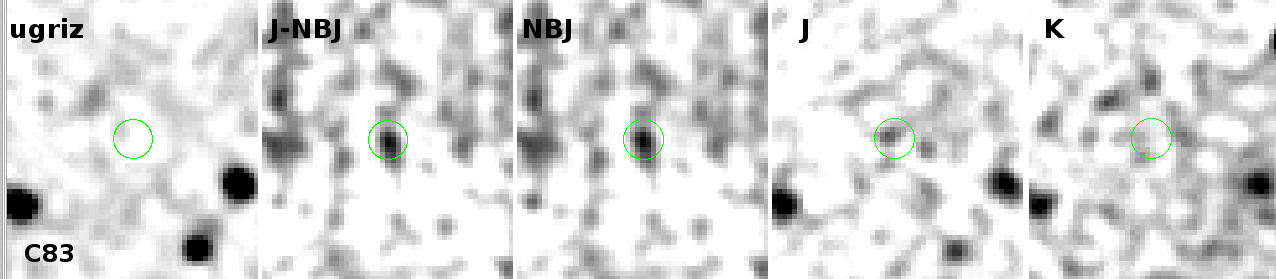}
\vspace{0.5ex}
\includegraphics[width=14cm]{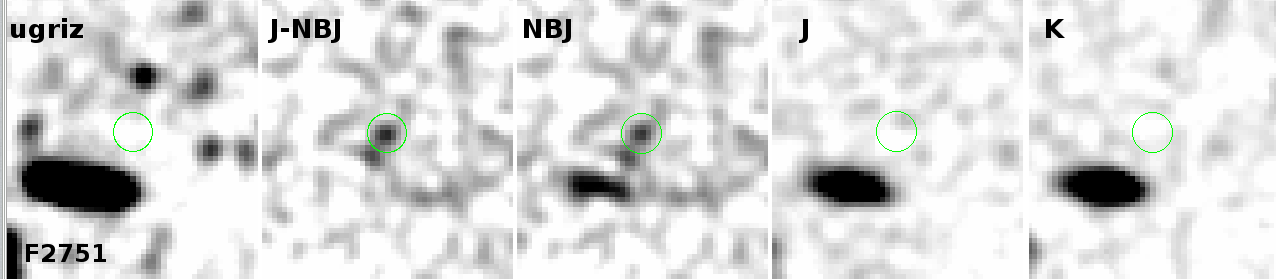}
\vspace{0.5ex}
\includegraphics[width=14cm]{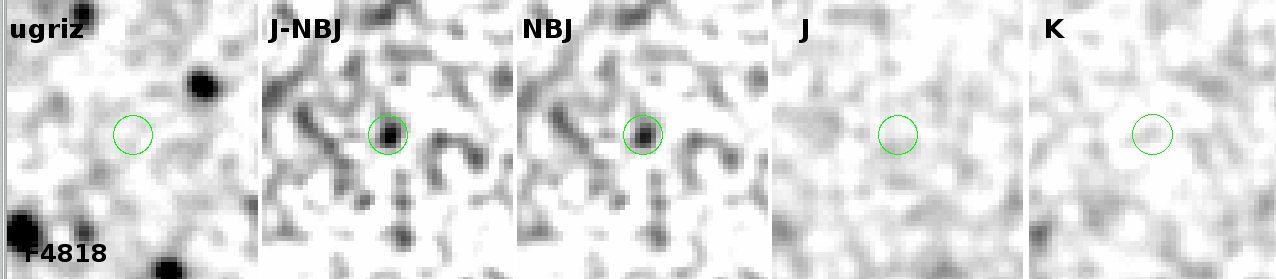}
\vspace{0.5ex}
\includegraphics[width=14cm]{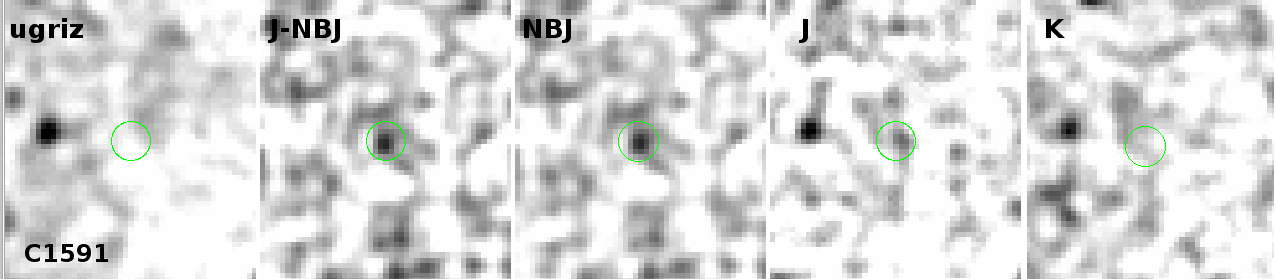}
\vspace{0.5ex}
\includegraphics[width=14cm]{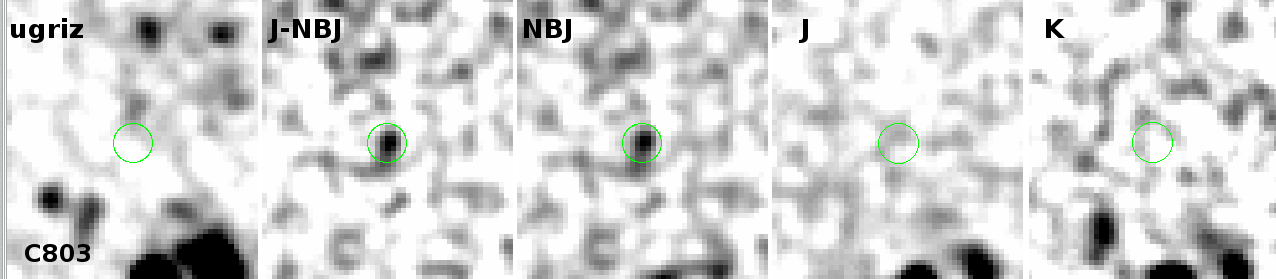}
\vspace{0.5ex}
\includegraphics[width=14cm]{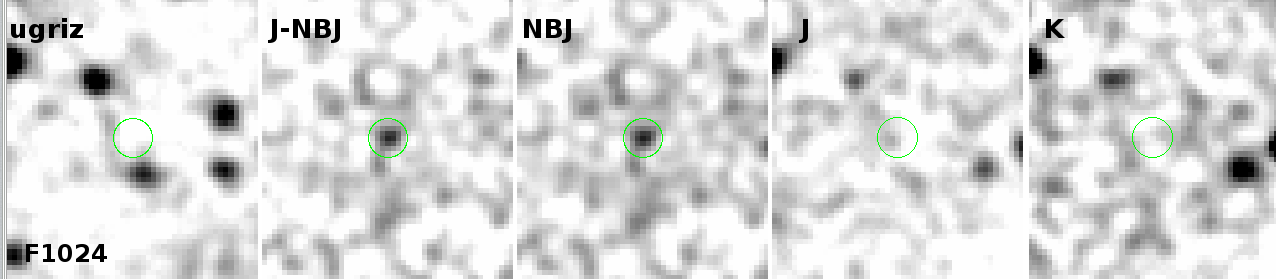}
\caption{\small{Thumbnails for the other seven Lyman-$\alpha$ candidates. Circles are placed at the center position of the thumb, corresponding to the position of the detection in the narrow-band. The angular scale of the thumbnails is $15\times15 ''$.}}
\end{figure*}